# X-ray photo-induced atomic motion in Phase Change Materials and conventional covalent chalcogenide glasses

I. Festi,[1,2,*] A. Cornet,[1] T. Fujita,[3] J. Moesggard,[3] A. Ronca,[1,4] J. Shen,[1] M. Sprung,[5] S. Wei,[3] F. Westermeier,[5] R. Escalier,[6] A. Piarristeguy,[6] G. Baldi,[2,†] and B. Ruta[1,‡]

[1]Institut Néel, Université Grenoble Alpes and Centre National de la Recherche Scientifique, 25 rue des Martyrs - BP 166, 38042, Grenoble cedex 9 France.

[2]Department of Physics, University of Trento, I-38123, Povo, Trento, Italy.

[3]Department of Chemistry, Aarhus University, 8000 Aarhus C, Denmark.

[4]European Synchrotron Radiation Facility, BP 220, F-38043 Grenoble, France.

[5]Deutsches Elektronen-Synchrotron DESY, Notkestrase 85, D-22607 Hamburg, Germany.

[6]ICGM, Université de Montpellier, CNRS, ENSCM, Montpellier, France.

X-ray Photon Correlation Spectroscopy (XPCS) enables direct access to atomic-scale dynamics in disordered materials, revealing both spontaneous and X-ray-induced relaxation processes. Here, we study two compositionally similar alloy glasses near their glass transition temperatures: the phase change material (PCM) $Ge_{15}Sb_{85}$ and the non-PCM alloy $Ge_{15}Te_{85}$. Both exhibit X-ray–induced atomic motion, yet with markedly different responses. $Ge_{15}Sb_{85}$ undergoes an immediate transition to a photo-induced yielding state, characterised by stationary dynamics governed solely by the absorbed dose. In contrast, $Ge_{15}Te_{85}$ shows a progressive slowing-down of the relaxation process, accompanied by a crossover from compressed to stretched exponential decay in the density autocorrelation functions. This behaviour is consistent with the emergence of liquid-like collective motion as supported by de Gennes narrowing in the wave-vector dependence of the dynamics at length scales comparable with the first sharp diffraction peak. Unlike $Ge_{15}Sb_{85}$, this alloy does not reach a stationary regime within experimental timescales, implying that the yielding transition occurs only after thousands of seconds with the available dose rate. Its response is also temperature dependent: at lower temperatures, the dynamics reflects intrinsic stress relaxation processes, whereas at higher temperatures becomes dose-controlled. These findings demonstrate that the dynamical response to X-ray excitation is not determined solely by chemical composition or bonding character, but results from the interplay between irradiation effects and structural relaxation pathways.

---

\* irene.festi@unitn.it

† giacomo.baldi@unitn.it

‡ beatrice.ruta@neel.cnrs.fr

# I. Introduction

X-rays are a fundamental tool for studying the intrinsic properties of a material [1]. Over the past decades, improvements in the brilliance and coherence of X-ray beams delivered in modern synchrotron sources have led to the development of powerful techniques to investigate the dynamical properties of matter, one of the most significant being X-ray Photon Correlation Spectroscopy (XPCS) [2]. This method provides information on the collective microscopic rearrangements that occur in disordered materials at nanometric and atomic scales [2]. In XPCS, the dynamics is investigated by analysing the temporal fluctuations in the intensity of speckles, which are interference patterns generated by coherent X-rays scattered from the material. By recording series of speckle patterns, XPCS captures these fluctuations, offering a direct measure of the particle dynamics. Studies on the atomic motion in hard materials have shown that the response of these systems to the high fluxes of X-ray beams employed in XPCS can vary significantly depending on the material family [3–10].

While in crystalline [3] and glassy [4] alloys XPCS provides information on the atomic diffusion and collective motion in the material, unexpected results have been reported in vitreous silica and germania [4]. In these materials, the incident flux was found to induce atomic dynamics within the samples, likely due to radiolysis [4]. This process occurs when an X-ray photon is absorbed, creating an electron-hole pair that leads to atomic displacement and the formation of a long-lived exciton. The latter gives rise to bond breaking and further atomic displacement. Subsequent studies have expanded this investigation to a variety of materials, including borate glasses [5–7, 10], sodium silicates [10], lithium-borates [6], and chalcogenide glasses [8, 11]. In particular, the studies on the oxide glass $LiBO_2$ [12] and the chalcogenide glass $GeSe_3$ [11] show that after a certain absorbed dose, the system reaches a stationary state of yielding, which is photo-induced by the X-rays. The specific absorbed dose depends on the physicochemical characteristics of the material.

In general, these studies collectively demonstrate that X-rays can be employed not only to probe spontaneous electron density fluctuations but also to drive them, granting access to material properties that would otherwise be inaccessible using conventional techniques. The response of a glass to X-ray irradiation contains information on the structure and the microscopic relaxation processes active in the material [4]. Hence, X-rays have emerged as a powerful probe to study spontaneous and induced dynamical fluctuations in complex systems in response to other physical parameters, such as temperature and pressure [11].

Despite extensive studies on metallic and oxide glasses, no XPCS studies have been performed on glassy phase change materials (PCM). These materials (such as $Ge_{15}Sb_{85}$, $Ge_2Sb_2Te_5$, GeTe, and Ag-In-doped $Sb_2Te$) are a class of functional materials relevant for electronic and optical switching devices [13, 14]. PCMs can be rapidly and reversibly switched between amorphous and crystalline phases by Joule heating or optical pulses. Their strong property contrast between the two phases can be used for encoding data or manipulating optical beams for functional devices. Although both PCMs and conventional chalcogenide-based non-PCM alloys (e.g. $Ge_{15}Te_{85}$, GeSe, and $GeSe_2$) are commonly considered as covalent glasses, they differ remarkably in physical properties such as liquid fragility, structural relaxations, crystallisation kinetics, and glass forming abilities [13, 14].

In this work, we characterize the evolution of the collective atomic motion induced by intense X-ray radiation in two compositionally similar glassy alloys, a PCM $Ge_{15}Sb_{85}$ and a non-PCM $Ge_{15}Te_{85}$, investigated in the vicinity of their glass transition temperature Tg. $Ge_{15}Sb_{85}$ is a metallic-like alloy classified as a phase change material (PCM) due to its rapid and reversible glass-crystal switching properties, which are fundamental for non-volatile memory applications [15–17]. $Ge_{15}Te_{85}$, on the other hand, is a conventional chalcogenide glass with high amorphous stability and slow crystallization kinetics, extensively studied in both liquid and glassy phases [16, 18–26]. The strong interest in this composition is driven by its favourable properties for selector devices, such as a high crystallisation temperature that ensures a wider operating range, a fast threshold switching speed enabling rapid off-to-on transitions, and a low threshold voltage for efficient switching [22].

From the dynamical point of view, recent studies performed by some of the authors showed that PCMs, including $Ge_{15}Sb_{85}$, shows a pronounced secondary β-relaxation in the glassy state, faster that the main α-relaxation process, implying the presence of local fast atomic motion and loosely bonded regions in the material [27]. By contrast, when Sb is replaced by Te, the similar composition $Ge_{15}Te_{85}$ shows virtually no β-relaxation, implying a rigid bonding network without local fast dynamics. The striking difference in relaxation behaviours is demonstrated to correlate with their markedly different glass forming abilities, thermal stabilities, and crystallisation kinetics [27]. While $Ge_{15}Te_{85}$ is a good glass former with a relatively stable supercooled liquid region, $Ge_{15}Sb_{85}$ can easily crystallize on approaching the glass transition temperature.

As we show below, this distinction also emerges from a different dynamical response to the X-ray beams. Specifically, $Ge_{15}Sb_{85}$ exhibits an immediate transition to a yielding state, similar to $GeSe_3$ [11], whereas $Ge_{15}Te_{85}$ is more resistant, requiring two to three orders of magnitude higher absorbed doses to reach the same state.

## II. Experimental methods
### A. Samples

Glassy samples of $Ge_{15}Sb_{85}$ and $Ge_{15}Te_{85}$ were prepared through magnetron sputtering deposition of stoichiometric targets synthesised following the same procedure as described in a previous study [16]. A thick uniform layer of samples was deposited onto a 4-inch Si (100) wafer using DC 10 W, 10 sccm Ar flow, and $5×10^{-3}$ mbar Ar pressure. A steep-edge profilometer determines the sputtering rates of the target compositions to estimate the sputtering time needed for the desired layer thickness $L$ (3-8 micrometers) to match the absorption length $\mu$ of the samples. The amorphous layers were separated from the Si substrate in the form of pieces or flakes, where single relatively large pieces (with a diameter of several millimetres) are selected for XPCS measurements. The nominal sample compositions were confirmed (within instrumental error ~ 3%) twice using x-ray fluorescence (XRF) with a Rigaku NEX CG, first for the sputtering target and then for the as-deposited layers. The main properties of the two samples are summarised in Table (I) [13, 21, 27].

|  | $T_g$ [K] | FSDP [nm$^{-1}$] | $\rho$ [g/cm$^3$] | $L$ [μm] | $\mu$ [μm] |
| --- | --- | --- | --- | --- | --- |
| $Ge_{15}Sb_{85}$ | 500 | 20.5 | 5.323 | 3 | 8.5 |
| $Ge_{15}Te_{85}$ | 412 | 19.6 | 5.53 | 8 | 6.9 |

Table I - Calorimetric glass transition temperature $T_g$, wavevector of the first sharp diffraction peak (FSDP) maximum, mass density $\rho$, sample thickness, and X-ray absorption length at the energies used in the XPCS experiments for $Ge_{15}Te_{85}$ and $Ge_{15}Sb_{85}$ [13, 21, 27].

### B. X-ray Photon Correlation Spectroscopy

XPCS experiments were carried out during two separate beamtimes at the beamline P10 of PETRA III synchrotron in Hamburg, Germany. In the first experiment, we measured the $Ge_{15}Te_{85}$ composition. An X-ray beam of 8 keV was focused onto a spot size of full width at half maximum (FWHM) of 2.8×2.2 µm² (H×V) with a photon flux of $\phi_0 \sim 4.5 \times 10^{10}$ ph/s, whereas in the second experiment, we measured the $Ge_{15}Sb_{85}$ using a beam energy of 8.1 keV, with a spot size of 3.4×2.0 µm² and a photon flux of $\phi_0 \sim 8 \times 10^{10}$ ph/s. The samples were placed under vacuum, to prevent oxidation, on a Cu sample holder, whose temperature was controlled by a resistive heater and a Lakeshore temperature controller. Measurements were performed at a wave vector $q$ corresponding to the position of the maximum of the FSDP for both compositions. $Ge_{15}Te_{85}$ was studied at temperatures T = [295, 325] K (corresponding to T $\simeq$ [0.7, 0.8]$T_g$, see values in Table I), while $Ge_{15}Sb_{85}$ was measured at T = [295, 350, 400] K (T $\simeq$ [0.6, 0.7, 0.8]$T_g$). Additionally, the $Ge_{15}Te_{85}$ composition was investigated as a function of the scattering vector q in a range [2.18; 27.37] nm$^{-1}$ at 295 K.

To explore the effect of the X-rays on the dynamics, the full beam intensity φ0 was attenuated through Si attenuators (att). The fractions of the transmitted beam intensity (i.e. the transmission $T_a$) and the relative beam fluxes $\phi$ used in the experiments are summarised in Table (II).

| | $Ge_{15}Sb_{85}$ | | | $Ge_{15}Te_{85}$ | |
|---|---|---|---|---|---|
| | $T_a$ | $\Phi$ [×10$^{10}$ ph/s] | | $T_a$ | $\Phi$ [×10$^{10}$ ph/s] |
| att = 0 | 100% | 8 | att = 0 | 100% | 4.5 |
| att = 2 | 45% | 3.58 | att = 2 | 45% | 1.96 |
| att = 4 | 22% | 1.73 | att = 4 | 22% | 1.26 |
| att = 6 | 10% | 0.77 | att = 6 | 10% | 0.91 |
| att = 8 | 5% | 0.4 | att = 8 | 5% | 0.39 |

Table II - Transmission percentages for the silicon attenuators used in the XPCS experiments for each material.

Each XPCS measurement was performed on a fresh sample spot to avoid any cumulative radiation effects. Different series of speckle patterns were collected in transmission geometry using an EIGER 4M detector positioned 1.8 m from the sample stage. Fluctuating speckle patterns are obtained when an amorphous system scatters a coherent X-ray, and they reflect the internal dynamics of the sample. Thus, the autocorrelation of the scattered intensity allows for the calculation of the microscopic dynamics, being

$$C_q(t_1, t_2) = \frac{\langle I_p(t_1) I_p(t_2) \rangle_p}{\langle I_p(t_1) \rangle_p \langle I_p(t_2) \rangle_p} \quad (1)$$

This function is called *two-times correlation function* (TTCF). $I_p(t)$ is the intensity measured at pixel $p$ and time $t$, and $\langle \cdots \rangle_p$ denotes an average over detector pixels at a given momentum transfer $q$. Examples of TTCFs are presented in Figures (2c-d). The main diagonal of the TTCFs corresponds to $t_1 = t_2$ and represents the correlation of a frame at a given time with itself, while all the off-diagonals are the correlation of a frame at time $t_1$ (or $t_2$) with a subsequent frame at time $t_2$ (or $t_1$).

From $C_q(t_1, t_2)$, the autocorrelation function $g_2(t)$ of the intensity fluctuations can be extracted:

$$g_2(q,t) = \langle C_q(t_0, t_0 + t)\rangle \tag{2}$$

where $t_0 = t_1$, $t = t_2 − t_1$, and $\langle \cdots \rangle$ represents the average over time. This means that the TTCFs are a map of the time evolution of the microscopic dynamics. The $g_2(q,t)$ provides information on the microscopic dynamics through the determination of the *intermediate scattering function* (ISF) $F(q,t)$ via the Siegert relation $g_2(q,t) = 1 + C_e|F(q,t)|^2$, where $C_e$ is the experimental contrast related to the degree of coherence of the X-rays [28]. The ISF is the time-dependent density-density correlation function and quantifies the extent to which the atomic configuration at a given moment resembles its configuration at a later time $t$.

III. Results

We model the $g_2(q,t)$ with a Kohlrausch-Williams-Watts (KWW) function [29]

$$g_2(q,t) = y_0 + Ce^{-2\left(\frac{t}{\tau}\right)^\beta} \tag{3}$$

where $C = C_e|f_q|^2$, with $f_q$ being the non-ergodicity factor and $C_e$ the experimental contrast, $\tau$ is the relaxation time, $\beta$ the degree of exponentiality of the decay, and $y_0$ accounts for the baseline [28]. All the parameters depend on the exchanged wavevector $q$, the temperature $T$, and the absorbed dose $D$. The total radiation dose absorbed by the sample during an exposure time $\Delta t$ is calculated as:

$$D = \frac{\epsilon}{m} F \Delta t \tag{4}$$

where $\epsilon$ is the X-ray energy and $F$ is the fraction of the incident beam, $\phi$, absorbed by the sample given by $F = \phi(1 − T_r)$, with $T_r$ being the sample transmission. As anticipated in the experimental section, the value of the incident flux $\phi$ was varied by attenuating the full beam $\phi_0$ through Si attenuators with transmissions Ta reported in Table (II). The incident fluxes employed in the experiments result in $\phi = \phi_0 T_a$. The dose also depends on the mass m within the scattering volume $V_s$: $m = \rho V_s$, where $V_s = AL$ (with $A$ as the focal spot area and $L$ the sample thickness) and $\rho$ is the mass density.

Consistent with previous studies on other chalcogenide glasses [8, 11], our data show that the probed atomic motion is also triggered by the X-ray irradiation in the studied compositions. This effect is observed at all investigated temperatures and at the probed q no appreciable structural damage is detected (see Figure (SM1)). However, within the limits of our experimental resolution, we cannot determine whether this apparent stability reflects a genuine resistance to structural damage, as reported for other glassy systems [4], or whether minor structural damage occurs below our detection threshold or in a different $q$ range.

An example of the probed dynamics is shown in Fig. (1) where we report auto-correlation functions measured with XPCS at different values of the X-ray flux at T = 0.7T$_g$, for both Ge$_{15}$Sb$_{85}$ (1a) and Ge$_{15}$Te$_{85}$ (1b). The long time-decay of the curves corresponds to the characteristic time associated with the probed dynamics. In both cases, this decay shifts to larger times with decreasing flux (from blue to green curves), indicating that the dynamics can be tuned by varying the intensity of the X-ray beam. The black dashed lines are the best fit to the experimental data according to equation (3). To ensure consistency between data collected at different X-ray fluxes, the comparison of the curves was performed by evaluating the $g_2(t)$ considering a fixed absorbed dose of 38 GGy for each composition.

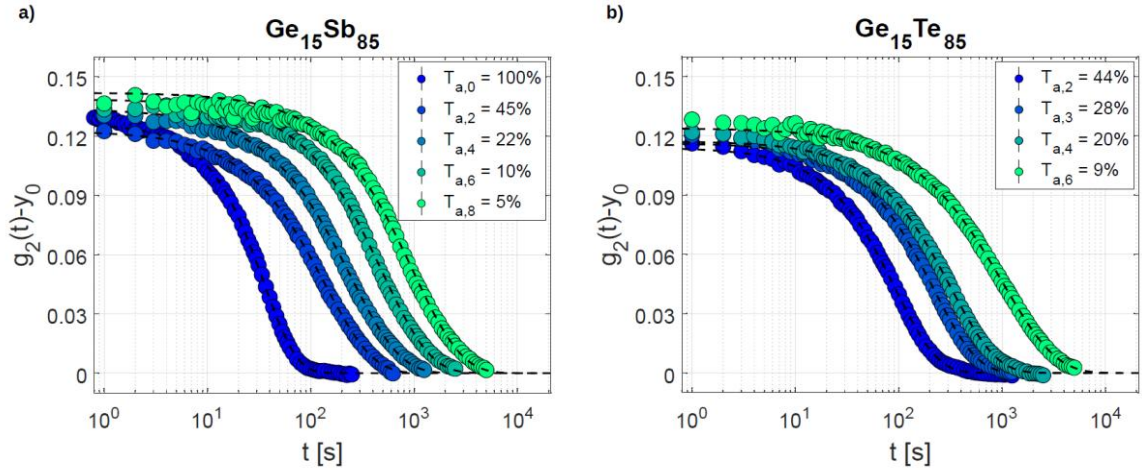

Figure 1 - Auto-correlation functions of the intensity fluctuations measured with XPCS at $0.7T_g$ in a) $Ge_{15}Sb_{85}$ and b) $Ge_{15}Te_{85}$ at varying beam attenuations. The curves of each sample are obtained by ensuring a total absorbed dose of 38 GGy for both samples. The dashed black lines represent the best fits to the KWW model. Each curve has been measured on a fresh sample position.

Although the dynamics of both glasses clearly depends on the X-ray flux, the two materials exhibit distinct responses to X-ray exposure. This is particularly evident from the temporal evolution of their relaxation processes at a fixed beam intensity as shown in Fig. (2). Here, the curves were evaluated by averaging the TTCFs according to Eq. (2) over time windows $\Delta t$, centred at waiting time $t_w = \Delta t(n + 1/2)$ with $n \in [0, 3]$, and $\Delta t$ = 900 s for the $Ge_{15}Sb_{85}$ and $\Delta t$ = 700 s for the $Ge_{15}Te_{85}$. In this context, the waiting time $t_w$ reflects the time elapsed since the onset of each measurement. As depicted in Fig. (2a), $g_2(q,t)$ curves acquired at different waiting times overlap each other, indicating that the dynamics of the $Ge_{15}Sb_{85}$ remains stationary over time. This is also evident from the TTCF, shown in Fig. (2c), where a constant intensity profile is observed along the main diagonal from the bottom left corner to the upper right corner. In contrast, $Ge_{15}Te_{85}$ exhibits a progressive slowing down of its dynamics over time. As shown in Fig. (2b), the $g_2(q,t)$ curves shift towards longer decay times as the waiting time progresses from 350 s to 2450 s, highlighting this effect. The presence of this process is also confirmed by the broadening of the intensity along the main diagonal of the TTCF of $Ge_{15}Te_{85}$ reported in Fig. (2d).

The slowing down of the dynamics is a spontaneous process in glasses held at temperatures close to $T_g$, known as physical aging [30]. As out-of-equilibrium systems, glasses naturally evolve over time towards more energetically stable configurations, leading to an increase in relaxation time. Under standard conditions, this aging process is thermally driven, as the system progressively explores its energy landscape. In our experiments, however, the observed slowing down is not solely spontaneous but is induced by the X-ray beam itself. The beam not only probes and stimulates atomic dynamics but also promotes a temporal evolution of the system. This effect may be linked to stress release occurring under irradiation or to local annealing processes initiated by the beam. These aspects will be discussed in greater detail later in this section.

The contrasting temporal evolutions of the dynamics in the two compositions imply a different response of the two systems to the accumulated dose $D$. To investigate this aspect, we express the relaxation time $\tau$, obtained from XPCS measurements and KWW fits (3), in terms of the absorbed dose $D_\tau$ required to induce the observed dynamics:

$$D_\tau = \frac{\epsilon}{m}\phi_0 T_a(1 - T_r)\tau \qquad (5)$$

This formulation accounts for the interaction between the material and the X-rays. In the case of a process purely induced by irradiation [4], the relaxation time τ is expected to scale inversely with the incident X-ray flux $\phi$, meaning that halving the beam flux ($T_a$ = 50%) doubles $\tau$. Consequently, the dynamics depend only on the absorbed dose, independent of flux variations. This expression also allows the comparison of the relaxation dynamics between different samples compositions and experimental parameters. To explore the distinct responses of $Ge_{15}Sb_{85}$ and $Ge_{15}Te_{85}$ to X-ray exposure, Figures (3) and (4) show the absorbed dose in relaxation time $D_\tau$ and the KWW exponent $\beta$ as functions of the accumulated dose $D$, for different attenuator transmissions $T_{a,i}$. This representation enables us to assess the dependence of the dynamical parameters on the absorbed dose and provides insight into the nature of the interaction, allowing us to distinguish between dynamics that are purely beam-induced and those influenced by additional mechanisms.

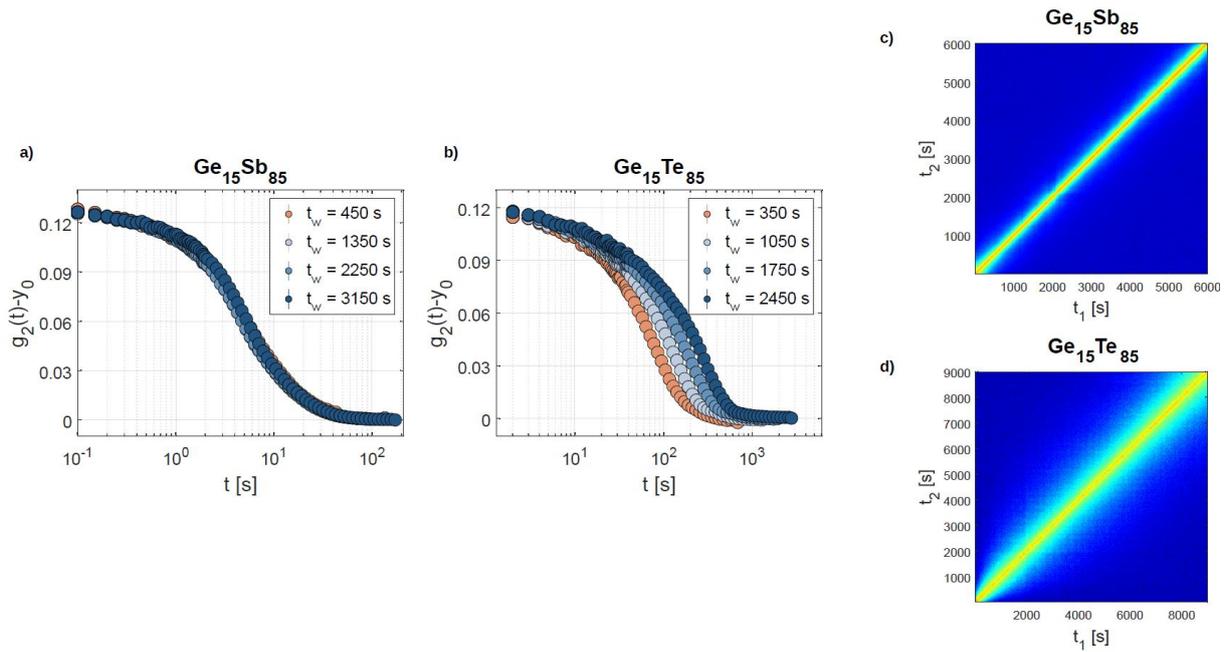

Figure 2 - Temporal evolution of the dynamics at 0.7$T_g$ for a fixed flux of the X-ray beam of 3.58×10$^{10}$ ph/s in $Ge_{15}Sb_{85}$ a) and of 1.96×10$^{10}$ ph/s in $Ge_{15}Te_{85}$ b). The dynamics remains stationary within the probed time window for the $Ge_{15}Sb_{85}$ a), while physical aging in the $Ge_{15}Te_{85}$ is signalled by the shift of the curves toward longer times with increasingly $t_w$ b). Two-times correlation functions measured in $Ge_{15}Sb_{85}$ c) and $Ge_{15}Te_{85}$ d) showing stable dynamics and aging, respectively.

Let's first focus on the dynamical behaviour of $Ge_{15}Sb_{85}$. The dependence of the XPCS parameters $\tau$ and $\beta$ on the accumulated dose $D$ is presented in Figure (3) at two temperatures, 0.7$T_g$ or 0.8$T_g$. Panels a) and c) show the dose $D_\tau$ accumulated in a relaxation time τ at the two temperatures, while the dose evolution of the $\beta$ parameter is plotted in panels b) and d). Different symbols and colours refer to different beam fluxes. The probed dynamics is beam-induced, since the $D_\tau$ parameter is independent of the beam flux for almost all the datasets, implying that the decorrelation time τ is inversely proportional to the beam flux. Overall, a good collapse of points corresponding to different beam fluxes is observed in all panels. There is, however, one dataset that follows a different trend. The blue circles at $T = 0.7T_g$, corresponding to full beam on the sample, exhibit a lower $D_\tau$ and a higher $\beta$ compared to the other points, likely due to local beam-induced heating that slightly increases the effective temperature of the probed spot. Apart from this dataset, the other points suggest that the system is in a yielded state at all the investigated doses, because the $D_\tau$ parameter is almost constant during the measurement and the stretching parameter $\beta$ is close to one, as in the case of $LiBO_2$ [12] and $GeSe_3$ [11].

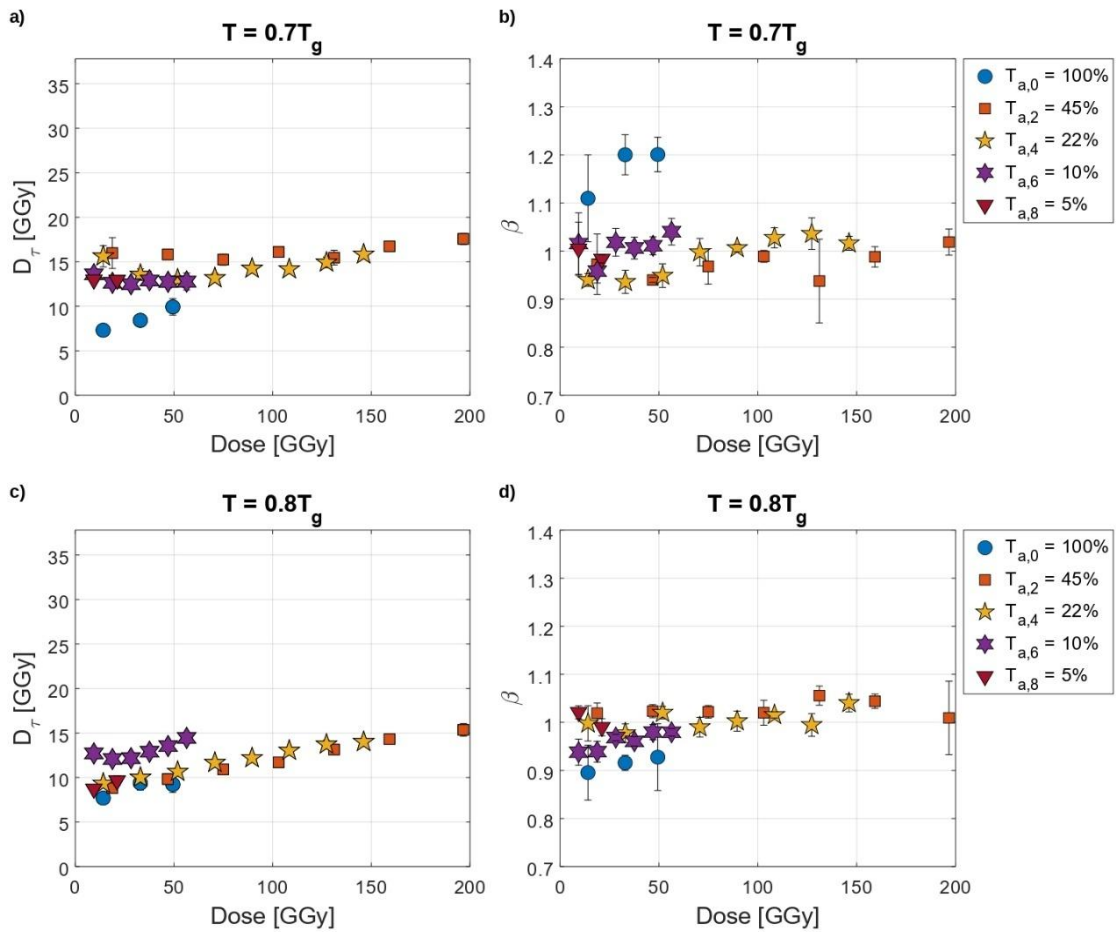

Figure 3 - Evolution of the dynamics in $Ge_{15}Sb_{85}$ as a function of the absorbed dose. a-b) At $0.7T_g$ and c-d) at $0.8T_g$, the $D_\tau$ and the $\beta$ tend to collapse into a master curve when plotted versus Dose. Both parameters are almost independent of the applied dose.

We now turn to the dynamical properties of $Ge_{15}Te_{85}$. As discussed before, a purely beam-induced dynamics is expected to scale with the accumulated absorbed dose $D$: if the dynamics is entirely governed by the X-rays, parameters such as the absorbed dose in XPCS relaxation times, $D_\tau$, and the shape parameter $\beta$ should collapse onto universal curves when plotted as a function of $D$. However, for $Ge_{15}Te_{85}$ at $0.7T_g$, this universal scaling is not observed (see Supplementary Figure (S2)). Although the beam induces the dynamics, $D_\tau$ and $\beta$ exhibit significant variations with the incident flux when plotted against $D$, suggesting that the absorbed dose alone does not fully describe the system's evolution.

To further clarify this behaviour, we plot $D_\tau$ and $\beta$ as functions of the experimental waiting time $t_w$, defined as the time elapsed since the start of exposure. Interestingly, when expressed in terms of $t_w$, the data collapse onto master curves across all fluxes (Fig. (4a-b)), revealing a universal, time-dependent evolution of the dynamics. To our knowledge, such a collapse with respect to the waiting time, rather than the absorbed dose, has not been previously reported in studies of beam-induced dynamics. This observation indicates that, while X-ray irradiation initiates atomic motion, the evolution at $0.7T_g$ remains governed primarily by intrinsic relaxation processes with timescales dictated by the material itself, rather than solely by the accumulated dose. This phenomenon can be rationalised by considering that $0.7T_g$ for $Ge_{15}Te_{85}$ corresponds to room temperature. It is therefore highly likely that the sample retains residual stresses from its preparation, which are progressively released under X-ray exposure. The stress relaxation drives the evolution of the dynamics, dominating the beam-induced effect.

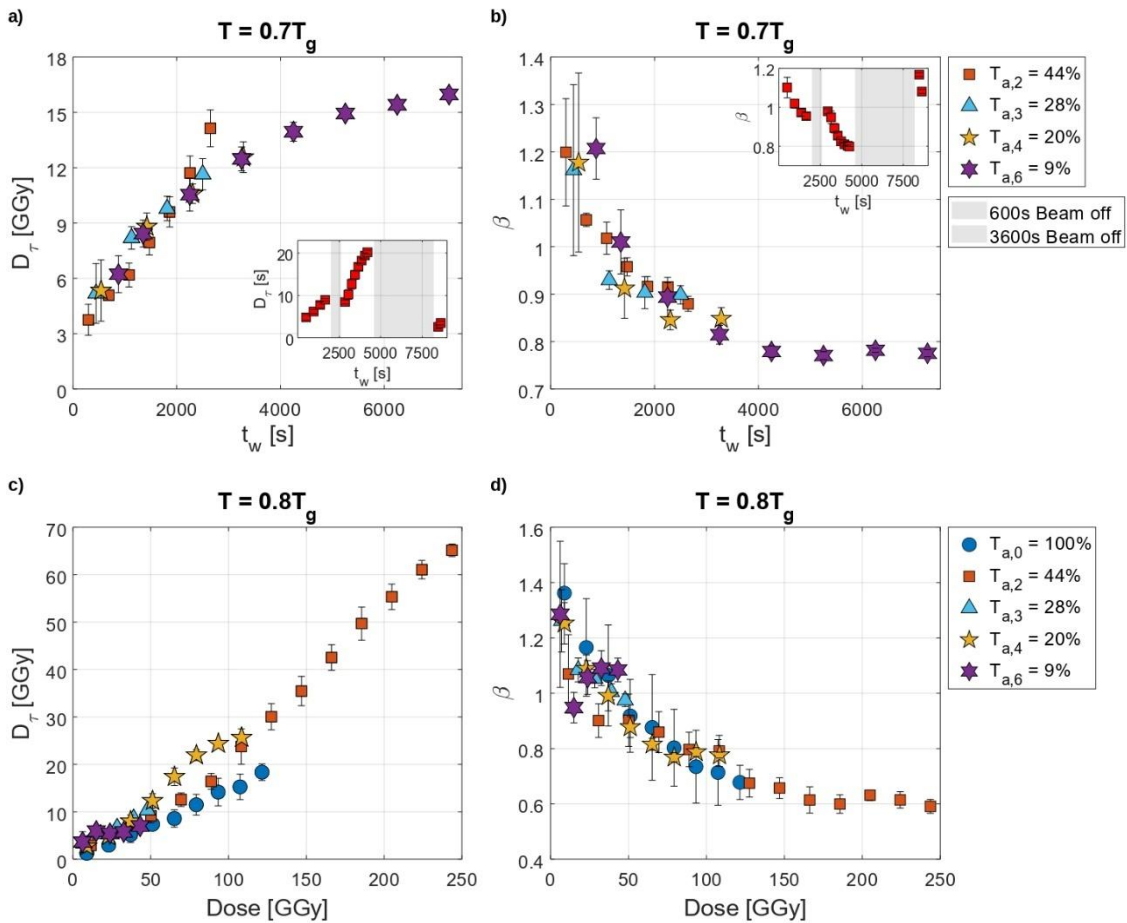

Figure 4 - a-b) Evolution of the dynamics in $Ge_{15}Te_{85}$ as a function of the waiting time $t_w$, at $0.7T_g$. $D_\tau$ and $\beta$ collapse into a master curve when plotted versus $t_w$, suggesting the dynamics is primarily influenced by intrinsic relaxations active in the material. While the beam does not irradiate the sample spot, both $D_\tau$ and $\beta$ tend to recover the initial state (insets of a-b)). c-d) At $0.8T_g$, $D_\tau$ and $\beta$ collapse in a master curve by plotting them versus the Dose. a-c) As the waiting time or the deposited dose increases, the dynamics slows down, suggesting a final plateau. b-d) With increasing the absorbed dose, the shape parameter $\beta$ decreases from $\beta > 1$, signature of compressed correlation functions, to $\beta < 1$, typical of stretched exponential decays.

Further evidence supporting this interpretation comes from "beam-of" experiments (Fig. (4) insets), where the X-ray beam was switched off for extended periods (see SM and Fig. (S3) for further details). For short interruptions (10 minutes), stress relaxation continues, reflecting intrinsic structural relaxation mechanisms triggered by the X-rays. Differently, for longer interruptions (1 hour), both $D_\tau$ and $\beta$ reverted toward their initial values, indicating a recovery of the original state in the absence of irradiation. Such a reversible response upon switching off the beam has never been observed in non-stationary beam-induced dynamics so far, and supports the absence of structural damage within the experimental doses.

This hypothesis of room temperature stress-release is confirmed by the fact that at $0.8T_g$, both $D_\tau$ and $\beta$ exhibit a universal scaling with the absorbed dose $D$, regardless of the incident flux (Fig. (4c-d)). This indicates that, at $0.8T_g$, the dynamics depends solely on the X-ray flux and is thus purely beam-induced. Remarkably, a modest 10% increase in temperature is sufficient to eliminate the stress-driven contributions observed at $0.7T_g$, placing the system entirely under the control of irradiation-driven dynamics.

Next, we examine how the dynamical parameters of the two compositions depend on the absorbed dose $D$ (or on the waiting time $t_w$) at both $0.7T_g$ and $0.8T_g$. To do that, we refer to Fig. (3) and (4) for $Ge_{15}Sb_{85}$ and $Ge_{15}Te_{85}$, respectively. As anticipated, the $Ge_{15}Sb_{85}$ glass exhibits stationary dynamics and a KWW exponent $\beta \lesssim 1$ at both investigated temperatures. These features correspond to a local state of photo-induced yielding by the X-rays, resembling the previous studies on other chalcogenide and oxide glasses [8, 11, 12]. In materials science, yielding refers to the transition from an elastic response, where deformation is reversible, to a plastic state, where the material undergoes irreversible atomic rearrangements. In glasses, yielding is particularly subtle since plastic events appear at very small scales and are associated with localised atomic rearrangements known as shear transformation zones [12]. Martinelli et. al. [12] showed that X-ray irradiation acts like an external force, generating tiny defects that trigger these hidden structural rearrangements, effectively pushing the material into a plastic, irreversible state at the microscopic scale. In the case of the chalcogenide glass $GeSe_3$, the yielding is induced within 100 s of exposure at ambient temperature [11], while the $Ge_{15}Sb_{85}$ composition reaches this stationary state within the exposure time of 1 s at the probed temperatures of 350 K and 400 K ($0.7T_g$ and $0.8T_g$, respectively). This difference in time scales is likely the consequence of the larger thermal fluctuations at the probed temperatures, which facilitate the occurrence of local yielding at a lower dose. In fact, in agreement with previous works, we also observe a transitory state with a weak aging at ambient temperature ($0.6T_g$) in Ge15Sb85 accompanied by a transition in the KWW parameter $\beta$ from compressed (≥1) to stretched (≤1) and the reaching of a stationary regime around 40 GGy (see (S4)). As discussed in detail in Ref. [12], these are the typical characteristics of the induced yielding transition.

On the other hand, the $Ge_{15}Te_{85}$ exhibits a strong dependence on the waiting time and absorbed dose at both temperatures, as evident from Figure (4). We see that neither the absorbed dose associated with the relaxation time $D_\tau$ nor the $\beta$ parameter reaches a complete plateau. Only the $D_\tau$ measured at $T_{a,6}$ and the $\beta$ at $T_a \geq 28\%$ at $0.7T_g$ suggest the proximity to a stationary regime. To understand whether this behaviour corresponds to the onset of the photo-induced yielding transition, we measured the wave-vector q dependence of the dynamics (and thus of the parameters $D_\tau$ and $\beta$) as a function of increasing absorbed dose (see Fig. (5)). Following Ref. [12], during the transition to local yielding, one should also observe a crossover from a $1/q$ wave-vector dependence of $\tau$ and an almost q-independent evolution of $\beta$, to a liquid-like behaviour.

At low $q$, a linear trend emerges in the log-log plot of $D_\tau$ versus $q$ (Figure 5a), indicating a power-law relationship between the relaxation time and the wavevector. By analysing this dependence in the range [2–7] nm$^{-1}$, we can access the nature of the particle dynamics and how it evolves with absorbed dose. Fitting the data with the expression $D_\tau(q) = (bq^\alpha)^{-1}$, where the exponent $\alpha$ characterizes the type of motion, we obtain $\alpha$ = 1.7±0.4, independently of dose (see inset of Figure 5a). This value is larger than the one reported for LiBO$_2$ glass, where $\alpha$ decreases from ∼ 1.1 to ∼ 0.3 with increasing dose. This trend was interpreted as a transition in the nature of collective motion, from ballistic-like dynamics to a regime where defect creation and annihilation are balanced, similar to a steady-state condition observed in liquids [12]. Differently, in the Ge15Te85, the larger value of $\alpha$ suggests the presence of diffusive motion at low $q$-s, with a negligible dose dependence. Additional data at lower $q$ values, well separated from structural contributions, are required to confirm this behaviour.

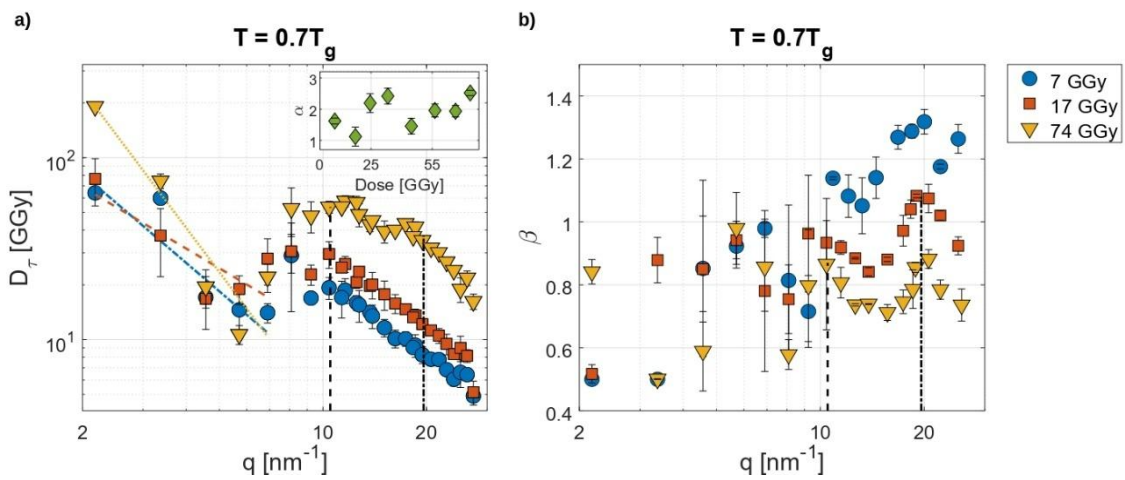

Figure 5 - Wave vector dependence of $\tau$ a) and $\beta$ b) at 0.7T$_g$ of Ge$_{15}$Te$_{85}$ for three selected doses: $D$ = 7 GGy (blue circles), $D$ = 17 GGy (red squares) and $D$ = 74 GGy (yellow triangles). The vertical dashed and dashed-dotted black lines represent, respectively, the position of the pre-peak and the main peak. a) The inset shows the dependence on the dose of the power-law exponent $\alpha$ describing the linear dependence of $\tau$ at low $q$-values.

At high $q$ (greater than 7 nm$^{-1}$), we observe an overall increase in $D_\tau$ and a concurrent decrease in $\beta$ with increasing dose $D$. At all doses, $D_\tau$ presents a peak at ∼ 10.5 nm$^{-1}$, which mirrors the corresponding prepeak of the static structure factor $S(q)$ [21]. At high doses, a peak arises at ∼ 19.6 nm$^{-1}$, corresponding to the main first sharp diffraction peak (FSDP). Even the KWW $\beta$ parameter exhibits features resembling the evolution of the $S(q)$, less evident at low doses but pronounced at higher doses. The resemblance between the q-dependence of τ and $\beta$, and that of the $S(q)$ is reminiscent of the phenomenon known as *de Gennes narrowing*, which takes the name from the narrowing at $q$-values corresponding to the FSDP of the width of the quasi-elastic contribution in neutron scattering experiments on liquids [31]. Given that the half-width at half maximum of the quasi-elastic peak is inversely proportional to structural relaxation time, $\Gamma(q) \propto 1/\tau$, this results in an increase in $\tau$ near structure factor peaks, implying that the most probable structural configuration is also the most stable. Therefore, the observed $q$-dependence in Ge$_{15}$Te$_{85}$ suggests that the sample volume irradiated by the X-rays displays a liquid-like behaviour and thus a local yielding as in the LiBO$_2$. This interpretation is supported by the transition of $\beta$ from values greater than 1, typical of a stress-driven dynamics of glasses well below T$_g$ [32], to values lower than 1, fingerprint of liquid-like heterogeneous relaxation dynamics [33, 34].

We note also that at low doses (below 0.5 GGy), the de Gennes narrowing is entirely absent for LiBO$_2$, in both $\tau$ and $\beta$ [12]. In our case, instead, the de Gennes narrowing effect is already present at the 10.5 nm$^{-1}$ pre-peak in $\tau$ ($D_\tau$), even in the lowest probed dose regime (below 7 GGy). In particular, at low doses only the signature of the pre-peak around ~ 10.5 nm$^{-1}$ is observed in the evolution of $D_\tau$ with $q$, while at high doses, the signature of the main peak around ~ 19.6 nm$^{-1}$ arises as well. Such a strong dynamical signature corresponding to the weak pre-peaks in structure factors is a general feature of glass-forming liquids, which has been associated with the influence of dynamic heterogeneity phenomena at the single and many particle level [35, 36].

## IV.     Discussion and conclusion

Our results reveal a strong correlation between the local atomic structure and the dynamical response to intense X-ray fluxes in two compositionally similar glasses, Ge$_{15}$Sb$_{85}$ and Ge$_{15}$Te$_{85}$. Consistent with previous findings in other oxides and chalcogenides [8, 11, 12], the X-rays can induce a local state of yielding in both compositions, where the amorphous solids present liquid-like features in the irradiated zone. The propensity of this phenomenon, however, strongly depends on the chemical and physical details of the material.

In the PCM Ge$_{15}$Sb$_{85}$, our data exhibit an almost instantaneous photo-induced yielding transition at 0.7T$_g$ and 0.8T$_g$. This rapid response is favoured by thermal motion, since at room temperature the transition requires ~ 400 s of continuous irradiation to occur (see SM). This is in agreement with what has been observed in GeSe$_3$ [11] at ambient temperature, where yielding occurs in ~ 100 s. The faster transition at room temperature in GeSe$_3$ is likely due to the Se being lighter than Sb, a fact supported by a faster relaxation time of $\tau$ ~ 2 s in GeSe$_3$ in comparison to $\tau$ ~ 100 s in Ge15Sb85 for similar doses. Increasing the temperature to 350 K (0.7Tg) enhances atomic mobility within Ge$_{15}$Sb$_{85}$, enabling the system to reach the yielding state almost immediately.

Unlike Ge$_{15}$Sb$_{85}$ and GeSe$_3$, Ge$_{15}$Te$_{85}$ does not completely reach the yielding state within the experimental timeframe, neither at 0.7T$_g$ (which corresponds to room temperature) nor 0.8T$_g$. This can be explained by the fact that Te atoms are heavier than Se or Sb atoms, reducing atomic mobility and resulting in a greater resistance to structural change upon irradiation with X-ray beams. This is consistent with the absence of visible structural changes on the FSDP, which are instead clear in GeSe$_3$. Moreover, Ge$_{15}$Te$_{85}$ exhibits recovery effects upon beam switching off: once the X-ray beam is turned off, the system begins to recover toward its initial dynamical state and the longer the absence of irradiation, the closer the dynamical parameters return to their original values (see insets of Fig. (4a-b)). In contrast, the GeSe$_3$, once it has reached the yielding state, shows a stable dynamics: during the absence of the beam, the system remains in that state. It is important to emphasise that these differing responses cannot be explained solely by considering the type of chemical bonding. Both GeSe$_3$ and Ge$_{15}$Te$_{85}$ are covalent glasses with good glass-forming ability and slow crystallisation kinetics. Nevertheless, they show distinct X-ray-induced dynamics, suggesting that factors beyond bonding, such as atomic mass, network topology, and local structural heterogeneity, play a crucial role in determining the onset and nature of yielding under irradiation.

The faster photo-induced atomic motion in the PCM Ge$_{15}$Sb$_{85}$ and the higher sensitivity to the X-rays induced yielding is likely the consequence of its locally loosely bonded microscopic structure and the presence of a pronounced secondary $\beta$-relaxation which promotes local mobility at the atomic scale [27]. This mobility is also confirmed by the fact that the dynamics does not scale completely with the dose $D$, strengthening the idea of the presence of an intrinsic dynamic process already active at the probed temperatures. By contrast, the rigid covalently bonded network in $Ge_{15}Te_{85}$ is less prone to changes of the local atomic environment, which reflects in longer relaxation times at the atomic scale and a pure beam-induced dynamics but at ambient temperature. At this temperature, the scaling law is not fulfilled, probably due to the presence of some additional stresses related to the sample preparation, which sum up with the irradiation-induced yielding transition.

Interestingly, we note that the $Ge_{15}Te_{85}$ exhibits a dependence on the absorbed dose similar to that reported in the oxide glass $LiBO_2$ [12]. In both systems, increasing the absorbed dose leads to a transition in the KWW $\beta$ shape parameter from a compressed ($\beta \geq 1$) to a stretched ($\beta \leq 1$) regime, which is accompanied by the emergence of a liquid-like wave-vector dependence of the dynamics at large doses. As previously discussed, this behaviour can be attributed to the transition from a purely beam-induced stress-driven motion to a local yielding transition of the material under X-ray irradiation [12].

It is, however, important to underline the main differences between $LiBO_2$ and $Ge_{15}Te_{85}$. One crucial factor influencing beam-induced dynamics is the X-ray absorption. $Ge_{15}Te_{85}$ consists of heavy atoms, which absorb X-rays more efficiently than the lighter elements present in $LiBO_2$. This difference is reflected in the experimental conditions: the thickness of the $Ge_{15}Te_{85}$ sample was comparable to its attenuation length at 8 keV (~ 8 μm), ensuring substantial absorption, whereas $LiBO_2$ was significantly thinner (125 μm) relative to its attenuation length at 8.4 keV (680 μm). Consequently, the dose range explored for $Ge_{15}Te_{85}$ reached tens of GGy, while for $LiBO_2$, it remained at a few GGy. Despite this large difference in the absorbed dose, both materials exhibit a similar evolution of the KWW $\beta$ parameter and τ(q) as a function of the absorbed dose. This suggests that atomic rearrangements in $Ge_{15}Te_{85}$ require a significantly higher number of electron-hole pairs, likely due to its heavier atomic composition. This interpretation aligns with the slower dynamics observed in $Ge_{15}Te_{85}$ (on the order of hundreds of seconds) compared to $LiBO_2$ (~ 20 s), despite its higher absorbed dose.

In conclusion, our XPCS study reveals distinct beam induced dynamical responses in glassy $Ge_{15}Sb_{85}$ and $Ge_{15}Te_{85}$ near their glass transition temperatures. Both systems are influenced by X-ray irradiation, yet their responses differ markedly despite their similar composition. $Ge_{15}Sb_{85}$ undergoes a rapid transition to a stationary yielding regime within the exposure time, characterized by a KWW parameter $\beta \lesssim 1$. By contrast, $Ge_{15}Te_{85}$ exhibits strong dose-dependent evolution, including aging and a progressive shift of the shape parameter β from compressed ($\beta > 1$) to stretched ($\beta < 1$). This transition is consistent with the emerging of liquid-like collective motion, as supported by de Gennes narrowing observed in the wavevector q-dependence of the dynamics. This glass does not reach a stationary regime within experimental timescales, indicating that local yielding requires thousands of seconds. This resistance to photoinduced yielding in $Ge_{15}Te_{85}$ is consistent with its heavier atomic constituents, which enhance structural rigidity and demand much higher deposited doses to induce local transformations. These findings align with its higher amorphous thermal stability and slow crystallisation kinetics [22], in contrast to $Ge_{15}Sb_{85}$ which exhibits rapid crystallisation kinetics and less covalent bonding [16]. Overall, neither chemical composition alone nor bonding character alone dictates the X-ray induced response of the non-PCM $Ge_{15}Te_{85}$ and PCM $Ge_{15}Sb_{85}$; instead, it emerges from the interplay between irradiation effects and intrinsic relaxation pathways. Finally, we note that such beam-induced dynamics may provide a novel means of probing β-relaxations in PCMs below Tg. Previous studies have shown that PCMs such as $Ge_2Sb_2Te_5$, GeTe, and AIST exhibit pronounced excess wings in the loss modulus associated with β-relaxations, in stark contrast to those chalcogenide non-PCMs whose excess wings are vanishingly small [37].

## V. Acknowledgements

We gratefully acknowledge PETRA III synchrotron (Hamburg, Germany) for providing beamtime, including experiments carried out at P10 under the LTP project II-20230020 EC. This project received funding from the European Research Council (ERC) under the European Union's Horizon 2020 research and innovation program (Grant Agreement No 948780).

## VI. Author Contributions

G.B. and B.R. conceived and designed the project. G.B., A.C., I.F., T.F., J.M., A.R., B.R., M.S., S.W., and F.W. performed the experiments. J.M., R.E., A.P. and R.E. carried out sample fabrication. I.F. analysed the data. G.B., B.R. and I.F. wrote the manuscript with contributions from all the authors. All the authors contributed to the discussion of the results, reviewed the data and provided feedback on the manuscript.

# Supplementary information to:

# X-ray photo-induced atomic motion in Phase Change Materials and conventional covalent chalcogenide glasses

I. Festi,[1,2] A. Cornet,[1] T. Fujita,[3] J. Moesggard,[3] A. Ronca,[1,4] J. Shen,[1] M. Sprung,[5] S. Wei,[3] F. Westermeier,[5] R. Escalier,[6] A. Piarristeguy,[6] G. Baldi,[2] and B. Ruta[1]

[1]*Institut Néel, Université Grenoble Alpes and Centre National de la Recherche Scientifique, 25 rue des Martyrs - BP 166, 38042, Grenoble cedex 9 France.*

[2]*Department of Physics, University of Trento, I-38123, Povo, Trento, Italy.*

[3]*Department of Chemistry, Aarhus University, 8000 Aarhus C, Denmark.*

[4]*European Synchrotron Radiation Facility, BP 220, F-38043 Grenoble, France.*

[5]*Deutsches Elektronen-Synchrotron DESY, Notkestrase 85, D-22607 Hamburg, Germany.*

[6]*ICGM, Université de Montpellier, CNRS, ENSCM, Montpellier, France.*

## I. Methods
### A. Absorbed dose calculation

The dose rate of X-ray absorption from a material is

$$D_{rate} = \frac{\epsilon e}{m} \phi_0 T_a (1 - T_r), \tag{1}$$

where $\epsilon$ and $\phi_0$ are respectively the energy and the flux of the X-rays, $e$ is the electron charge, and $T_a$ is the transmission of the Si attenuators employed to attenuate the beam intensity. The dose rate also depends on the absorbed fraction of the beam

$$1 - T_r = 1 - e^{-\frac{L}{\mu}}, \tag{2}$$

where $L$ is the sample thickness, while $\mu$ is the absorption length of the material at the energy $\epsilon$. Finally, the dose rate depends on the mass $m$ within the scattering volume $V_s$, according to $m = \rho V_s$, being $\rho$ the mass density of the sample. The volume is found considering an elliptical beam cross section, whose semiaxes correspond to the FWHM/√2ln2 values.

The absorbed dose is found considering the exposure amount of time $\Delta t$ of the sample spot to the beam

$$D = D_{rate} \Delta t \tag{3}$$

### B. Data treatment of $\tau$ and $\beta$ vs $q$

The $q$-dependence of the dynamics of $Ge_{15}Te_{85}$ shown in Figure (5) of the main article was measured at $0.7T_g$ with attenuator 2. Due to the lower statistics at high $q$-values, we applied a two-point averaging to the $\beta$ parameter to improve data clarity and reduce statistical fluctuations. This averaging was not applied to the $\tau$ relaxation time, as it is a significantly more robust parameter and did not require smoothing.

## II. Beam damage

The scattered intensity from the samples showed no significant variation over the duration of the XPCS measurements (Figure (S1)), suggesting that either the underlying structure remained largely unchanged by the X-ray beam, indicating no appreciable structural damage, or that any structural modifications occurring were not detectable through the analysis of the mean scattered intensity alone. This observation implies that the observed dynamical changes are likely not a direct consequence of gross structural degradation of the amorphous network within the probed timeframe.

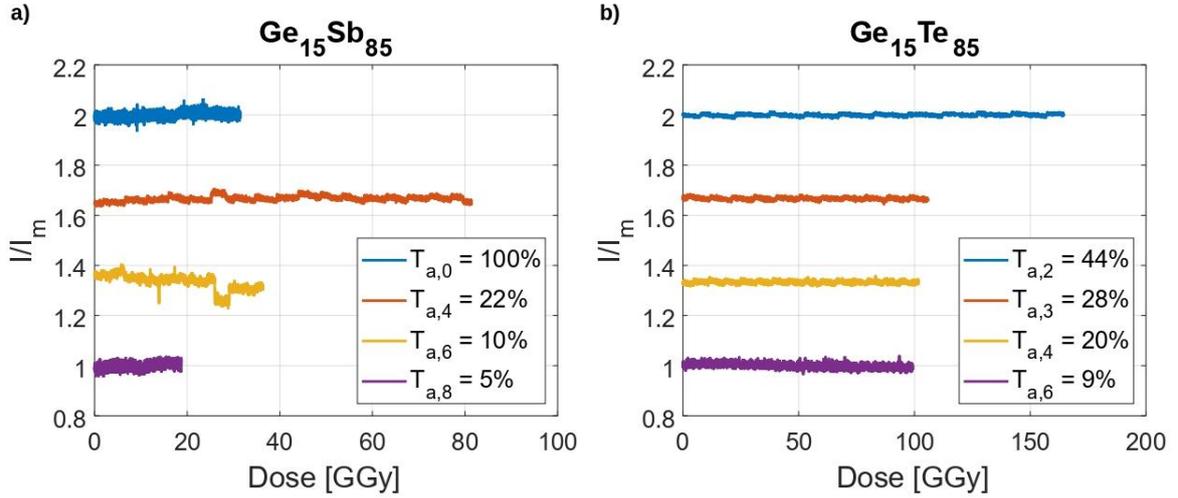

Figure S1 - The average scattered intensity collected by the detector for each frame, normalized to the relative beam intensity at various attenuation levels. As time passes, no appreciable structural damage is occurring.

## III. $Ge_{15}Te_{85}$
### A. $T=0.7T_g$ versus absorbed dose $D$

At $0.7T_g$, if we plot $D_\tau$ and $\beta$ versus the absorbed dose $D$ we observe a failure in the collapse into a universal master curve, differently from what was observed at $0.8T_g$.

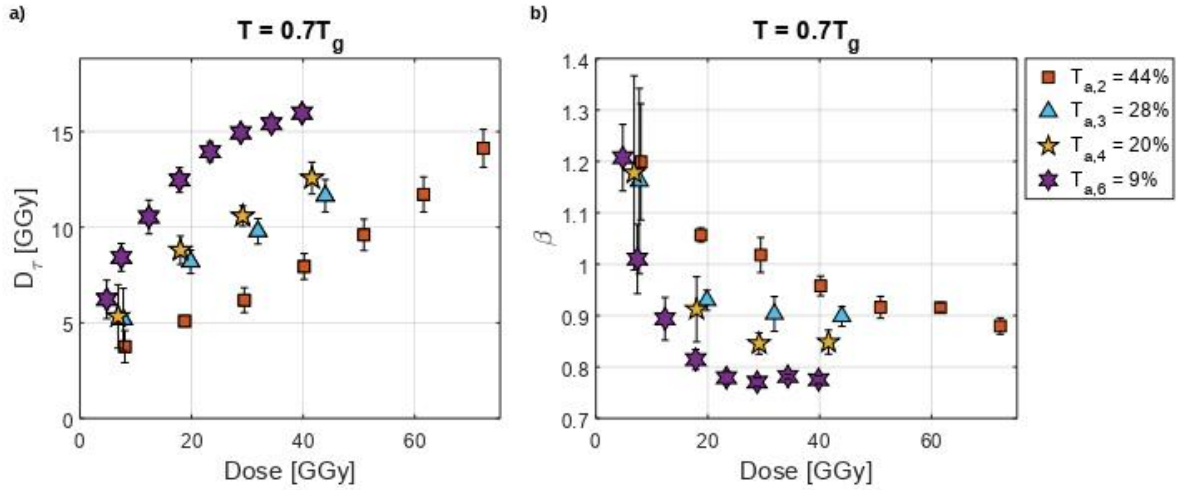

Figure S2 - Dynamics evolution at T = 0.7$T_g$ of $Ge_{15}Te_{85}$. a) $D_\tau$ and b) $\beta$ fail to collapse into a master curve when plotted versus $D$.

### B. Recovery of dynamics after beam interruption

To further investigate the nature of the beam-induced dynamics in $Ge_{15}Te_{85}$, we conducted experiments where the X-ray beam was switched off after an initial measurement on a single sample spot, and then turned back on for a subsequent measurement on the same location. Our findings revealed that during the period when the beam was off, the dynamics of the material tended to revert towards its initial state (Figures (S3)).

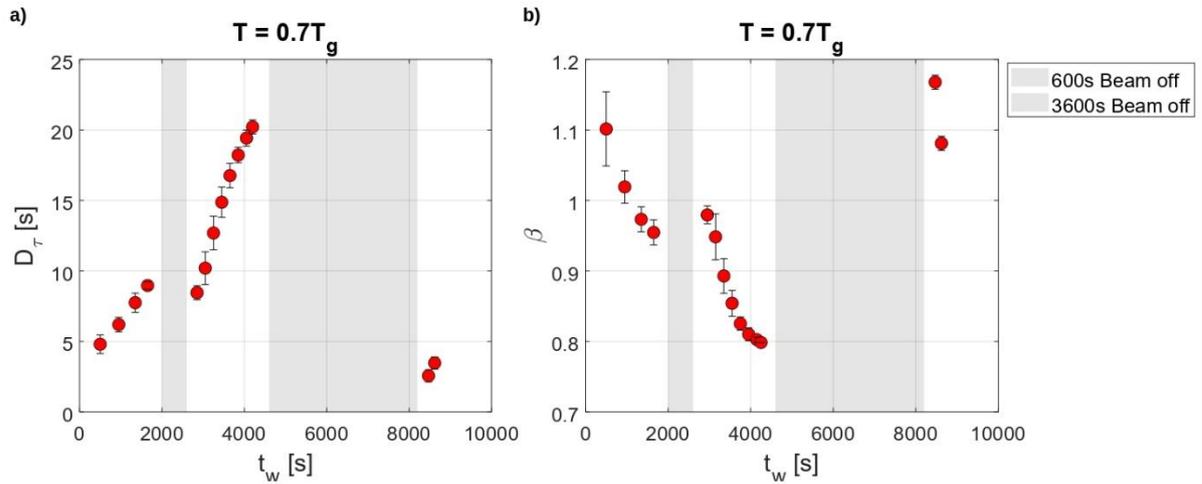

Figure S3 - Dynamics evolution at T = 0.7$T_g$ of $Ge_{15}Te_{85}$ switching off the beam for 600 s and for 3600 s. a) The relaxation time $\tau$ increases in time, but while the beam is off it tends to go back to the initial values. b) the shape parameter $\beta$ decreases in time, but while the beam is off it tends to go back to the initial values.

### IV. $Ge_{15}Sb_{85}$
#### A. T=0.6$T_g$

While $Ge_{15}Sb_{85}$ generally exhibits stationary dynamics at 0.7$T_g$ and 0.8$T_g$, measurements at 0.6$T_g$ reveal some hints of aging, specifically at the highest beam intensities. At lower fluxes, the relaxation time $\tau$ remains stable. At 0.6$T_g$, we also observe a decrease in the stretching exponent $\beta$ from approximately 1.1 to 0.9, a trend reminiscent of $Ge_{15}Te_{85}$ but significantly less pronounced

(where $\beta$ shifts from ~ 1.3 to ~ 0.6). This transient behaviour in $Ge_{15}Sb_{85}$ at room temperature ($0.6T_g$) is likely due to residual stresses introduced during sample preparation. These stresses may make the material more susceptible to the X-ray beam, a phenomenon far more pronounced in $Ge_{15}Te_{85}$, before dissipating at higher temperatures where the intrinsic stability of the glass dominates.

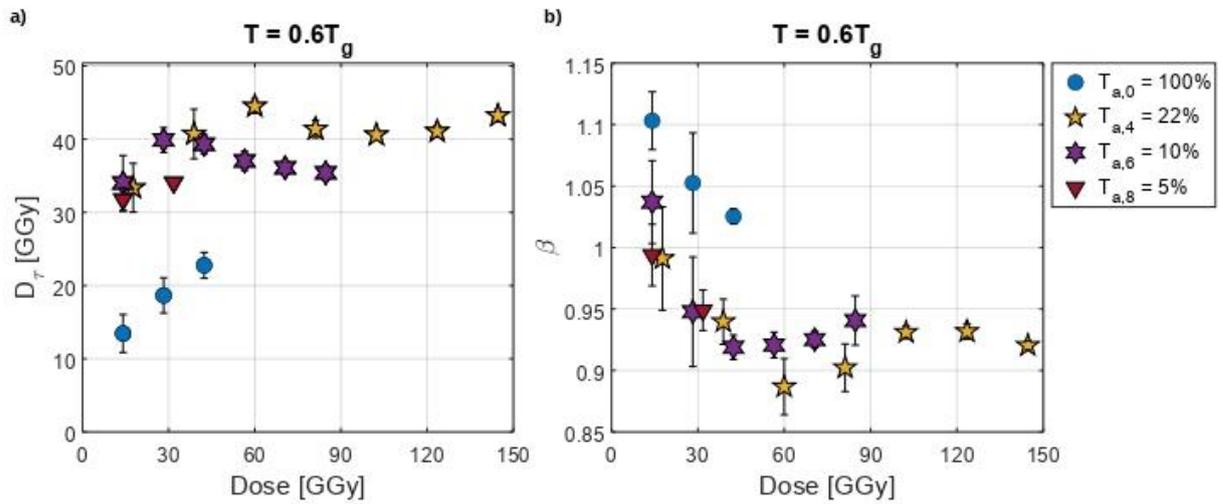

Figure 6 - Dynamics evolution at T = $0.6T_g$ of $Ge_{15}Sb_{85}$. a) the relaxation time $\tau$ shows a constant behaviour, except for very high beam intensities. b) $\beta$ decreases from ~ 1.1 to ~ 0.9.